\documentclass[10pt]{article}
\usepackage[french]{babel}
\usepackage{epsfig,shadow}
\usepackage{fancyhdr}
\usepackage{hyperref}
\usepackage[T1]{fontenc}
\usepackage[latin1]{inputenc}
\usepackage{graphicx}
\usepackage{amsfonts}
\usepackage{color}
\usepackage{amsmath}
\usepackage{amssymb}
\usepackage{float}
\usepackage{algorithm}
\usepackage{algorithmic}

\def\keywords{\vspace{.5em}
{\textit{Keywords}:\,\relax%
}}
\def\endkeywords{\par}

\begin{document}

\title{Intégration des intergiciels de grilles de PC\\ dans le nuage SlapOS : le cas de BOINC}

\author{Christophe Cérin, Nicolas Greneche, Alain Takoudjou\thanks{Nous remercions en particulier Jean-Paul Smets et Cédric de Saint-Matin de Nexedi pour les échanges fructueux au cours de la rédaction de cet article.}\\
 \\
Université Paris 13, PRES Sorbonne Paris Cité\\
LIPN UMR CNRS 7030\\
99, avenue Jean-Baptiste Clément\\
F-93430 Villetaneuse, France\\
alain.takoudjou@lipn.univ-paris13.fr}

\date{12 novembre 2012}

\maketitle

\begin{abstract}
  Dans cet article nous exposons les problèmes et les solutions liés à
  l'intégration des intergiciels de grilles de PC dans un cloud, en
  l'occurrence le cloud libre SlapOS. Nous nous concentrons sur les
  problématiques de recettage qui décrivent l'intégration ainsi que la
  problématique du confinement d'exécution qui constituent deux
  aspects systèmes des architectures orientées service et du \og Cloud
  Computing\fg. Ces deux problématiques dénotent par rapport à ce qui
  se fait traditionnellement dans les clouds parce que nous ne nous
  appuyons pas sur des machines virtuelles, qu'il n'y a pas de centre
  de données (au sens du cloud). Par ailleurs nous montrons qu'à
  partir du modèle de déploiement initial nous pouvons prendre en
  compte non seulement des applications Web, B2B\ldots mais aussi des
  applications grilles ; ici un intergiciel de grilles de PC qui
  constitue une étude de cas.  
\end{abstract}

	
\section{Introduction}

Le cloud computing ou \og informatique dans le nuage\fg désigne une
forme de traitement et un ensemble de ressources informatiques
massivement extensibles, exploités par de multiples clients externes
sous forme de services. Plus précisément selon le NIST (National
Institute of Standards and Technology), le cloud computing est l'accès
via le réseau, à la demande et en libre-service à des ressources
informatiques virtualisées et mutualisées
\cite{NIST:Publication800.145}. La \og cloudification\fg,
c.à.d. l'intégration d'une application dans un cloud doit s'envisager
sous les angles :

\begin{itemize}

\item Software as a Service (SaaS) : l'application est découpée en services ;

\item Data as a Service (DaaS) : les données sont disponibles sur le réseau ;

\item Platform as a Service (PaaS) : la plate-forme est granulaire ;

\item Infrastructure as a Service (IaaS) : l'infrastructure est virtualisée.

\end{itemize}

C'est en 2009 qu'a eu lieu une réelle explosion du cloud sur le marché
avec l'arrivée de Google App Engine \cite{APPEngine}, Microsoft Azure
\cite{WindowsAzure}, IBM Smart Business Service \cite{IBM-SBS} et bien
d'autres.  En mars 2010, IELO, Mandriva, Nexedi et Tiolive qui sont
quatre éditeurs de logiciels libres se réunissent pour fonder la \og
Free Cloud Alliance\fg (FCA) dans le but de promouvoir des Clouds Open
Source \footnote{Free Cloud Alliance:
  http://www.freecloudalliance.org/press/fca-Press.Contact/news-free-cloud-alliance},
elle propose une offre globale réunissant IaaS, PaaS et SaaS,
constituée de tous les composants libres nécessaires aux applications
Progiciel de Gestion Intégré (ERP), gestion de la relation client
(CRM) ou gestion de la connaissance (KM).

En 2011 le projet Resilience, retenu dans le cadre du 12\`eme appel à
projets du fonds unique interministériel (FUI-12) fut lancé. Ce projet
regroupe des universitaires (Paris 13, Institut Télécom, INRIA) et des
industriels dont Nexedi, Sagem Morpho, Wallix, Vifib. Ce projet
comporte une partie \og prospective amont\fg qui est relatée dans cet
article.

Le projet Resilience se base sur le cloud libre SlapOS
\cite{DBLP:conf/IEEEscc/Smets-SolanesCC11}, dont les principales
  caractéristiques sont les suivantes, en première approximation :

\begin{itemize}

\item il ne repose pas sur de la virtualisation ;

\item il ne repose pas sur des centres de données ;

\item il reprend en partie une architecture de grille de PC
  \cite{cerin} : des machines à la maison abritent des services et des
  données ; un \og maître\fg contient un annuaire des services et les
  services eux mêmes ;

\item la propriété d'interopérabilité s'obtient \og à la Grid'5000\fg
  c.à.d en déployant le démon SLAPGRID sur des noeuds d'Amazon,
  d'Azure\ldots puis en installant sur ces n\oe uds les bonnes
  versions logicielles. Cela évite d'utiliser par exemple Libcloud qui
  est une librairie servant à l'intéropérabilité des clouds\ldots mais
  rien ne garantit que chacun des clouds tourne la m\^eme version de
  Libcloud ce qui est généralement une condition nécessaire au bon
  fonctionnement global.

\end{itemize}

Le problème général que nous adressons est le suivant : comment faire
en sorte qu'un service de Grille de PC soit déployable depuis
n'importe o\`u, n'importe quand (à la demande) et sur n'importe quels
dispositifs (smartphones, tablettes, PC\ldots). Une première option
est d'intégrer tel quel dans le nuage les intergiciels les plus
utilisés à ce jour comme BOINC et Condor. C'est l'objet de cette
étude. La deuxième option serait de repenser les interactions entre
les composants usuels d'une grille de PC en terme de paradigme issus
du Web 2.0. C'est l'objet d'un autre travail que nous ne commentons
pas plus ici.

Dans cet article, nous illustrons nos contributions à partir de notre
expérience de \og cloudification\fg de l'intergiciel de grille de PC
BOINC dans SlapOS ce qui fait ressortir les points clés architecturaux
de SlapOS, les principaux concepts liés aux systèmes d'exploitation
qui sont mis en jeux. Les difficultés résident dans le fait que BOINC
est organisé différemment par rapport aux applications Web qui sont le
\og fond de commerce\fg de SlapOS. La question est donc de savoir si
le modèle SlapOS est suffisamment général pour \og cloudifier\fg aussi
des intergiciels de grilles. Nous ouvrons également une discussion sur
le modèle de confinement d'exécution et sur son couplage avec
SlapOS. 

Ainsi, dans cet article nous présentons une méthodologie d'intégration
de BOINC dans le nuage SlapOS ainsi qu'une première réflexion quant au
confinement d'exécution dans SlapOS. Si l'on veut qu'un utilisateur
accepte de voir tourner une partie d'une application sur son propre
smartphone ou tablette, il est nécessaire de lui garantir que
l'application ne pourra pas être intrusive. Ce problème n'est pas
propre aux grilles de PC, mais il est d'importance et sa solution est
directement liée à la manière d'intégrer une application, ici
BOINC. L'état actuel de développement de SlapOS n'inclut pas de
solution pour ce problème.

L'organisation de l'article est donc le suivant. Dans le paragraphe
\ref{one} nous introduisons les concepts clé de SlapOS, les problèmes
traités et notre positionnement. Le paragraphe \ref{two} décrit les
expériences et ce que nous en retirons en terme d'organisation future
de SlapOS. Le paragraphe \ref{trois} conclut cet article.

\section{Enjeux scientifiques et techniques}\label{one}

\subsection{Présentation générale de SlapOS}

SlapOS (Simple Language for Accounting and Provisioning Operating
System \cite{DBLP:conf/IEEEscc/Smets-SolanesCC11}) est un système
d'exploitation distribué et Open Source qui fournit un environnement
permettant d'automatiser le déploiement des applications, tout en
incluant des services de comptabilité et la facturation gr\^ace à l'ERP5
(Progiciel de Gestion Intégré libre basé sur la plate-forme Zope et le
langage Python). Basé sur la devise que \og tout est processus\fg,
SlapOS combine le Grid Computing (en particulier BonjourGrid
\cite{bonjourgridTWO,bonjourgridTHREE}) et les techniques des ERP pour
fournir des composants IaaS, PaaS et SaaS, tout cela à travers un
démon appelé SLAPGRID. Les forces de SlapOS sont la compatibilité avec
tout système d'exploitation, particulièrement les systèmes GNU Linux ;
la compatibilité avec toutes les technologies logicielles, le support
de plusieurs infrastructures (IaaS) différentes et plus de 500
recettes disponibles pour les applications grand public telles que
LAMP (Linux Apache MySQL PHP).

La cloudification d'intergiciels de Grid Computing ne faisait donc pas
partie du c\oe ur du métier des personnes ayant introduit la solution
SlapOS. Cet article démontre que BOINC peut avantageusement s'intégrer
au cloud SlapOS et nous montrons les efforts nécessaires tout en
soulignant les concepts clés qui rendent cette intégration
aisée. L'objectif est donc d'étudier et de montrer que l'on peut, \og
en un clic\fg, instancier, configurer des familles de logiciels et les
déployer sur Internet.

\subsubsection{Concepts clés de SlapOS}

L'architecture de SlapOS est constituée de deux types de composants :
SlapOS Master et des SlapOS Nodes. Le SlapOS Master indique au SlapOS
Node quel logiciel doit être installé et aussi quelle instance d'un
logiciel spécifique sera déployé ; il agit comme un annuaire
centralisé de tous les SlapOS Nodes ; il connaît l'emplacement o\`u les
logiciels sont situés ainsi que tous les logiciels qui y sont
installés. En se basant sur la date de démarrage, la date d'arrêt (ce
qui représente un intervalle d'exécution) et aussi sur les ressources
utilisés par une application, le Master peut faire la facturation pour
chaque utilisateur. Le SlapOS Node peut être hébergé sur un ou
plusieurs datacenters ou encore sur un ordinateur local, il communique
avec le SlapOS Master à l'aide du SLAP Protocol et renseigne de façon
périodique sur l'état des ressources dont il dispose. Le rôle du
SlapOS Node est d'installer et d'exécuter des processus tandis qu'un
Master a pour rôle d'allouer les processus aux SlapOS Nodes.

SlapOS Node s'exécute sur un noyau appelé \og SlapOS kernel\fg, il est
constitué d'une distribution minimale du système GNU Linux, d'un démon
nommé SLAPGRID, d'un environnement d'amorçage des applications
(technologie Buildout) et de Supervisord qui sert à contrôler les
processus en cours d'exécution.

\medskip

Pour introduire les concepts clés, nous discutons, de manière
générale, de l'installation d'un nouveau logiciel dans SlapOS. Ainsi
SLAPGRID reçoit du Master une requête pour installer un logiciel,
celui-ci télécharge alors le fichier de description du logiciel que
l'on appelle \og Buildout profile\fg et lance le processus d'amorçage
Buildout qui va installer le logiciel. Buildout est un système de
compilation développé en python et utilisé pour créer, assembler et
déployer des applications composées de plusieurs pièces dont certaines
peuvent être non-basé sur Python. Il permet de créer une configuration
Buildout et de l'utiliser pour reproduire/cloner le logiciel plus tard. Il
est capable d'exécuter un programme C, C++, ruby, java, perl, etc. Il
joue un rôle que l'on peut rapprocher à celui de GNU Make.

SLAPGRID peut aussi recevoir du Master une requête demandant de
déployer une instance d'un logiciel, il utilise alors \verb+Buildout+
pour créer tous les fichiers (fichiers de configurations et programmes
à exécuter) nécessaires, à l'aide de Supervisord lance le démarrage de
l'application.

Un logiciel sur un SlapOS Node est appelé \og Software Release\fg et
il est constitué de l'ensemble des programmes et composants
nécessaires à son fonctionnement. A partir d'un Software Release, on
peut créer plusieurs instances du logiciel correspondant qu'on appelle
\og Software Instance\fg. Le concept de Software Instance renvoie à
l'idée selon laquelle un serveur peut exécuter de façon indépendante
un nombre élevé de processus d'un même logiciel. Puisque ces processus
utilisent la même mémoire partagée, l'empreinte mémoire est surchargée
et permet d'exécuter une autre instance avec des ressources mémoires
minimales, contrairement au principe de virtualisation. Il est donc
possible d'exécuter sur un SlapOS Node plus de 200 instances d'un même
logiciel. Dans le cas de Tiolive Services, 200 ERP5 peuvent être
exécuté sur un CPU dual core standard.

%

\medskip

Une partition SlapOS ou \og Computer Partition\fg peut être vue
comme un conteneur léger ou un enceinte clos \cite{ComputerPartition}, 
il fournit un niveau d'isolation raisonnable (inférieur à celui d'une 
machine virtuelle) basé sur la gestion des utilisateurs et des groupes 
par le système d'exploitation hôte. Chaque Partition est constituée 
d'une adresse IPv6, une adresse IPv4 locale (privée), une interface 
TAP \cite{Virtualisation.Systeme:TAB} dédié nommée \og slaptapN\fg, 
un nom d'utilisateur de la forme \og slapuserN\fg et un répertoire 
dédié (généralement /srv/slapgrid/slappartN).  SlapOS est configuré 
pour fonctionner avec IPv6, l'un des avantages étant la possibilité 
d'avoir un très grand nombre de partitions. 

Une {\it partition} est destinée à contenir une seule application,
celle-ci peut être une Software Instance quelconque ; la Software
Instance est alors accessible depuis l'adresse IP de la {\it
  partition}. Seul l'utilisateur slapuserN a les droits de lecture et
d'écriture sur les données de la partition $N$, de même les processus de
la partition sont exécutés et contrôlés par l'utilisateur
slapuserN. Toutes ces règles permettent d'assurer une certaine
sécurité et évite des accès non autorisés venant par exemple des
processus d'une autre partition. Puisque SlapOS s'exécute en mode
administrateur (root) il a donc la possibilité d'accéder et de
paramétrer la partition N sans avoir besoin d'être slapuserN.

%

\medskip

Généralement, SlapOS regroupe les environnements d'exécution des
logiciels sous forme de {\it stack} (pile). Le concept de {\it stack}
dans SlapOS représente un environnement de base pour le déploiement
d'une classe d'application bien précise. Nous pouvons prendre
l'exemple des applications Web basées sur PHP, une {\it stack}
(actuellement nommée LAMP) permettra alors d'installer l'ensemble des
composants nécessaires pour l'exécution de ces applications (MySQL,
PHP, Apache, etc). La {\it recette} associée permettra de démarrer les
services. L'objectif ici est de fournir une méthode généralisant le
déploiement des applications de même type tout en simplifiant la
création de leurs profils Buildout. Toute la complexité est renvoyée
au niveau de la {\it stack}. Dans le même ordre d'idée, les composants
(Apache, PHP, etc) sont portés de manière séparée et indépendante dans
SlapOS ce qui permet d'utiliser le même composant dans plusieurs
{\it stacks} ou Software Release.


\medskip

Le fonctionnement de SlapOS repose aussi autour des technologies
externes Buildout et Supervisord. Afin d'automatiser le déploiement
des applications, SlapOS a recours à la technologie Buildout en
exploitant les concepts de {\it parts} et de {\it recette}. Une {\it
  part} représente tout simplement un objet, un paquet python ou un
programme manipulé par Buildout. La {\it part} est référencée par son
nom et sera installée dans le répertoire de l'application à laquelle
elle est associée. Pour chaque {\it part} est définie une {\it
  recette} qui contient sa logique de gestion ainsi que les données
qui serons utilisées. Une {\it recette} est un objet qui sait comment
installer, mettre à jour ou désinstaller une {\it part} précise. Dans le
cadre de SlapOS on dispose d'un annuaire de {\it recettes} nommé
\emph{slapos.cookbook} qui contient près de 105 recettes pour le
déploiement d'applications et de composants. Le paquet
slapos.cookbook peut être consulté sur
\emph{http://pypi.python.org/pypi/slapos.cookbook/}

\subsection{Confinement d'exécution, sûreté de fonctionnement}

La première idée de conception de SlapOS est de considérer que tout
est processus ainsi, le système est basé sur la collection d'un
ensemble de processus qui communiquent entre eux à base de services
internet utilisant des protocoles de communications. Les processus
sont surveillés par l'outil Supervisord qui permet de les relancer en
cas de problèmes. Nous n'allons pas plus loin ici sur ces questions de
tolérance aux pannes car cela est en dehors du cadre mais ces
préoccupations sont abordées dans le projet Resilience. SlapOS combine
des hébergements non fiables, des serveurs dédiés à moindre coût et
des clouds à domicile pour atteindre une fiabilité de 99,999\%. SlapOS
est en réalité beaucoup plus fiable que les approches traditionnelles
du cloud puisque, en sélectionnant des sources indépendantes, il peut
survivre en cas de force majeure : grève, tremblement de terre,
coupures d'électricité, etc.

\section{Description des expériences}\label{two}

\subsection{Introduction}

Supposons que vous souhaitiez qu'une de vos machines s'intègre dans 
SlapOS. La démarche générique est la suivante :

\begin{enumerate}
\item télécharger et/ou installer une image SlapOS ;
\item créer un compte utilisateur sur le master (http://vifib.net);
\item amorcer cette image (configuration automatique du réseau, 
formatage du disque local si besoin c.à.d. si on souhaite installer 
les partitions SlapOS sur ce disque) ;
\item lancer la commande \verb+slapprepare+ afin de configurer les partitions 
SlapOS, enregistrer le n{\oe}ud au niveau du master (le login et le 
mot de passe du compte utilisateur précèdent sont requis) et démarrer 
le démon SLAPGRID ce qui autorise la communication entre le n{\oe}ud 
et le master ;
\item la dernière opération consiste à demander à l'administrateur du
  master l'enregistrement du/des applications à installer sur le n\oe
  ud. Il faut fournir une url de la forme http://git.erp5.org/gitweb/
  slapos.git/blob/refs/heads/grid-computing:/software/boinc/software.cfg.
\end{enumerate}

\subsection{Articulation des fichiers de recettage}

\begin{figure}[httb]
  \includegraphics[width=15cm]{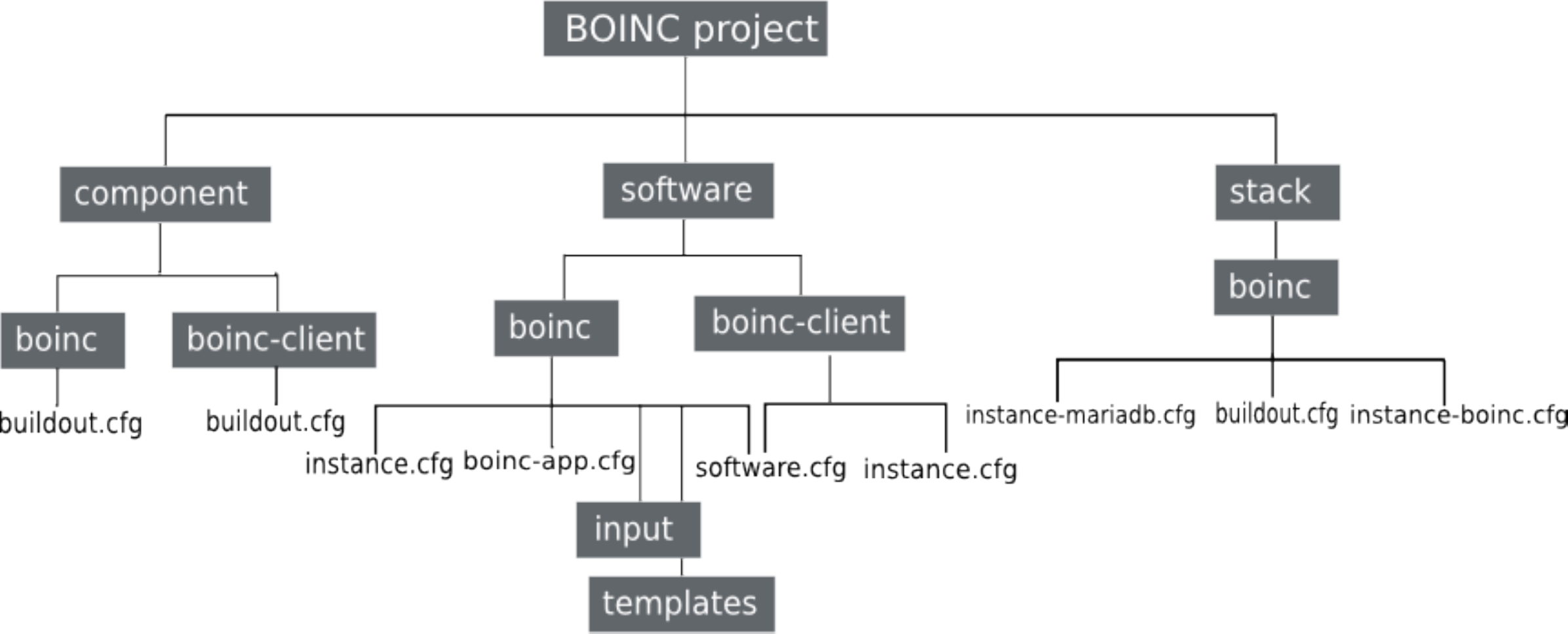}
  \caption{Présentation de l'arborescence du projet BOINC pour SlapOS.}
  \label{f1}
\end{figure}

L'intégration des applications à SlapOS passe par l'écriture des 
profils Buildout, constitués principalement du fichier \verb+software.cfg+, 
qui va ensuite faire la référence à tous les autres fichiers requis. 
Dans le cadre de l'intégration des applications de type BOINC (applications 
qui tournent avec BOINC) nous avons conçu l'architecture présentée
à la figure \ref{f1}. 

BOINC est une plate-forme de calcul distribué. Pour l'utiliser, on
peut créer un projet puis fournir une ou plusieurs applications, des
fichiers de données et les fichiers de configuration pour l'exécution
de l'application sur sa plate-forme. Il convient donc de créer une
{\it stack} que nous avons nommée BOINC et qui va permettre
d'installer de configurer un environnement d'exécution pour les
applications BOINC dans SlapOS. Ainsi la création d'un projet et
l'intégration d'une ou plusieurs applications à exécuter se réduira
simplement à la définition des paramètres permettant de personnaliser
notre projet et au téléchargement des fichiers utiles. Cet aspect nous
semble un effort raisonnable. 

BOINC est subdivisé en trois applications distinctes que nous
discutons maintenant.

\subsubsection{BOINC Server}
L'installation du serveur est plus complexe. Sur la Figure \ref{f1},
nous avons appelé tout simplement cette installation \og BOINC \fg,
elle est donc constitué de la stack nommée \og BOINC \fg, d'un
composant nommée \og BOINC \fg et d'une application exemple nommée
aussi \og BOINC \fg.

BOINC est une application avec de nombreuses dépendances pour sa
compilation, toutefois, la plupart de ses composants sont déjà
disponible dans SlapOS. Le profil du composant BOINC permet donc de
compiler BOINC dans SlapOS en y intégrant tous les autres composants
nécessaires pour cette compilation. Notons sur la Figure \ref{f1}
aussi l'utilisation des options \emph{--disable-manager} et
\emph{--disable-client} qui permettent de désactiver la compilation du
client et du manager.

La stack BOINC intervient à deux niveaux ici, pendant la compilation 
et pendant le déploiement. La compilation fait référence au fichier 
buildout.cfg de la stack, tandis que le déploiement de la stack fait 
référence au fichier instance-mariadb.cfg (pour installer MySQL dans 
une partition) et instance-boinc (pour installer BOINC, Apache, PHP, 
SVN, etc dans une autre partition). Elle est générique c'est-à dire 
qu'elle peut être utilisée pour l'installation ou le déploiement de 
plusieurs projets et applications BOINC différentes sans aucune 
modification supplémentaire, il suffira de lui passer les 
configurations appropriées.

Le software BOINC est une application exemple appelée à être modifiée
pour adapter l'application BOINC qu'on souhaite installer. Il est
constitué du fichier software.cfg qui est utilisé pendant la
compilation et permet de créer un software Release BOINC contenant la
stack et le composant BOINC et de télécharger le binaire et les
fichiers de/des applications à installer. Le dossier contient l'objet
input qui contient les données pour des applications à
exécuter. Toutefois ces fichiers peuvent se situer sur un emplacement
ou un serveur distant, ils seront alors téléchargés pendant la
compilation. Le dossier templates contient les templates
d'entrées/sorties à utiliser pour la configuration du WORKUNIT de
l'application, il peut être personnalisé.

Pour le déploiement de BOINC Server, nous avons ajouté à slapos.cookbook 
deux recettes :
\begin{itemize}
\item La recette \emph{slapos.cookbook:boinc} qui permet de déployer 
un projet BOINC vide ; elle est utilisée directement dans la stack.
\item La recette \emph{slapos.cookbook:boinc-app} qui permet de déployer 
une application dans une instance BOINC existante. Il est donc possible 
de déployer plusieurs applications pour un même projet, en appelant 
successivement la même recette \emph{slapos.cookbook:boinc-app} 
pour les différentes applications.
\end{itemize}

Nous présentons maintenant des échantillons de fichiers de
configuration de parts Buildout permettant de compiler et de
déployer un projet BOINC muni de l'application exemple
\emph{upper\_case}, qui met en majuscule un texte fourni.  La part
\emph{boinc-application} (qui se trouve dans le fichier software.cfg)
télécharge le binaire et collecte les paramètres du WORKUNIT qui sera
crée, elle est utilisée pendant la compilation. Sa forme est la
suivante :
\begin{verbatim}
1.  #Download Boinc Application Binary and configure project
2.  [boinc-application]
3.  recipe = hexagonit.recipe.download
4.  url = ${boinc:location}/libexec/examples/upper_case
5.  download-only = true
6.  filename = upper_case
7.  #Application configuration
8.  app-name = upper_case
9.  version = 1.0
10. exec-extension = 
12. platform = x86_64-pc-linux-gnu
13. #Work Unit: wu-name without blanc space: wu-number number of work unit
14. wu-name = simpletest
15. wu-number = 1
\end{verbatim}
Dans cette description on remarque que nous spécifions, à la ligne 3,
la recette qui permet de télécharger le binaire spécifié via l'URL de
la ligne 4. Les autres lignes concernent le type de plate-forme requis
(ligne 12) et le nom du WORKUNIT (ligne 14) ainsi que le nombre de WORKUNIT
(ligne 15) à affecter à l'application spécifiée à la ligne 8.

La part \emph{boinc-app} présentée ci-dessous (et contenue dans le
fichier boinc-app.cfg) est utilisée pendant le déploiement ; elle
hérite (ligne 3) de BOINC Server contenu dans la stack et est basée
sur la recette slapos.cookbook:boinc-app (ligne 4). Elle utilise les
fichiers et paramètres définis dans le fichier software.cfg et plus
particulièrement ceux de la part \emph{boinc-application}. Entre les
lignes 6 et 17 il ne s'agit que d'un transfert des paramètres définis
dans le fichier software.cfg.

\begin{verbatim}
1.  #Deploy a Boinc application in existing boinc server instance.
2.  [boinc-app]
3.  <= boinc-server
4.  recipe = slapos.cookbook:boinc.app
5.  binary = ${boinc-application:location}/${boinc-application:filename}
6.  #app-name should be unique (for all app deployed in a boinc instance)
7.  app-name = ${boinc-application:app-name}
8.  version = ${boinc-application:version}
9.  platform = ${boinc-application:platform}
10. extension = ${boinc-application:exec-extension}
11. dash = ${dash:location}/bin/dash
11. #templates
12. template-result = ${template_result:location}/${template_result:filename}
13. template-wu = ${template_wu:location}/${template_wu:filename}
14. #Work Unit
15. wu-name = ${boinc-application:wu-name}
16. wu-number = ${boinc-application:wu-number}
17. input-file = ${template_input:location}/${template_input:filename}
\end{verbatim}

Il suffira de mettre plusieurs parts de ce type dans le fichier
boinc-app.cfg pour déployer autant d'applications dans le serveur,
tout en prenant soin de donner des noms différents à ces parts ainsi
qu'aux options \emph{app-name}, \emph{wu-name}.  Les paramètres de
personnalisation du serveur BOINC sont passés directement pendant le
déploiement de l'instance à travers les paramètres d'instance appelés
\emph{slapparameter}. Il s'agira du nom du projet, du nom de la base
de données, etc.

\subsubsection{BOINC Client}

Le client représente un n{\oe}ud de calcul, il sera donc possible
d'utiliser SlapOS pour avoir des ressources supplémentaires pour les
calculs. Sur la figure \ref{f1}, nous avons appelé cette partie
boinc-client, elle est constituée d'un composant nommée \og
boinc-client \fg et de son application nommée aussi \og boinc-client
\fg. Nous avons ajouté à slapos.cookbook la recette
\emph{slapos.cookbook:boinc-client} qui permet de déployer une
instance d'un BOINC Client. L'url et la clef du compte utilisateur
crée sur le serveur BOINC doivent être fournies à l'instance à travers
les paramètres d'instance appelés slapparameters pour permettre la
connexion entre le serveur et le client BOINC.

\subsubsection{BOINC Manager}
Nous n'avons pas intégré cette application à SlapOS car étant entièrement 
graphique elle doit être installé sur un ordinateur personnel, c'est une 
application de bureau.

\subsection{Vers un nouveau mécanisme de confinement d'exécution dans SlapOS}

\subsubsection{Utilisation du MAC (Mandatory Access Control)}

Le MAC est une forme de contrôle d'accès différent du traditionnel DAC
(Discretionary Access Control). Dans les systèmes DAC, les permissions
sur les objets sont laissées à la discrétion du possesseur. La
philosophie du MAC est d'appliquer une politique de sécurité sur un
système à laquelle même le super utilisateur (root) doit se plier. La
base d'une politique MAC est le Type Enforcement (TE)
\cite{badger1995practical}. Dans le TE, les objets du système sont \og
typés\fg et on définit les interactions permises entre ces différents
types. Au dessus du TE, il est possible d'implémenter des modèles de
politiques de sécurité type MLS (MultiLevel
Security)\cite{bell1973secure}. En pratique, le TE
délimite les interactions systèmes potentielles d'un programme (même
si il s'exécute sous l'identité du super utilisateur). Nous allons
utiliser SELinux \cite{smalley2001implementing} qui propose une
implémentation du TE intégrée au noyau Linux.

\subsubsection{SELinux}

SELinux est une implémentation de TE basée sur l'architecture FLASK
\cite{peter2001integrating}. Les entités du système sont décomposées
en sujets et objets. Les sujets sont les entités \og actives\fg (par
exemple les processus). Les objets sont les entités \og passives\fg
(par exemple les fichiers). Toutes ces entités se voient attribuées un
label. Une politique de sécurité définit les interactions permises
entre un label de type sujet et un autre de type objet. La granularité
de ces interactions est au niveau de l'appel système (lecture,
écriture, exécution, transition). Une politique peut être appliquée en
mode \og targeted\fg, c'est à dire sur certains services seulement
(essentiellement ceux exposés sur le réseau) ou \og strict\fg, c'est à
dire sur tous les services.

\subsubsection{Pourquoi SELinux ?}

SlapOS propose de déployer des services à l'intérieur de
partitions. Du point de vue de la sécurité, une partition est composée
d'un ou plusieurs processus et des fichiers associés (configuration,
librairies, logs etc.). Le tout est déployé dans un répertoire
particulier du système. Le problème est qu'une intrusion sur une
partition (suite par exemple à une faille sur un service, un mot de
passe faible ou une erreur de configuration) suivie d'une escalade de
privilèges compromet toutes les partitions du système (une fois les
privilèges administrateur obtenus, toutes les partitions de la
machines physique sont accessibles). Ce type de faille n'est pas
propre à SlapOS mais à tous les systèmes basés sur le DAC
\cite{loscocco1998inevitability}. Les systèmes de Cloud traditionnels
contournent le problème en ajoutant une couche de virtualisation qui
(entre autre chose) fait office de mécanisme de confinement. Cependant
cette couche supplémentaire augmente la surface d'attaque du système.
GrSEC est un autre outil de contrôle d'accès mandataire comparable à
SELinux mais SELinux est intégré au noyau Linux et il est plus \og
supporté\fg que GrSEC. Nos avons donc choisi de travailler avec
SELinux.

\subsubsection{Intégration de SELinux dans SlapOS}

L'idée est d'utiliser SELinux pour :
\begin{itemize}
\item limiter les interactions des partitions avec le système de base ;
\item Interdire les interactions entre partitions.
\end{itemize} 

L'expérimentation pourra se faire en deux temps. Une première phase
serait de faire tourner SELinux en mode \og targeted\fg sur les
partitions. Chaque partition opère avec une identité dédiée et exécute
des binaires situés sur le système de base. Prenons comme exemple une
partition ayant le rôle de serveur et tournant sous l'identité web01,
une politique générale pourrait être :
  \begin{itemize}
    \item web01 peut exécuter le binaire httpd ;
    \item web01 peut lire sa configuration ainsi que les pages qu'il
      doit servir dans la partition dédiée (objets de type fichiers
      situés dans le répertoire de la partition) ;
    \item web01 peut écrire ses logs dans la partition.
  \end{itemize}

  Toutes les autres interactions sont interdites pour les processus
  tournant sous l'identité web01. L'étape suivante est de passer la
  politique en mode \og strict\fg pour protéger non plus seulement les
  services déployés par SlapOS mais le service de déploiement lui
  même. En conclusion, la démarche expérimentale est donc :

\begin{enumerate}
\item Intégrer la phase de labellisation du système de fichier lors de
  l'instanciation du disque local pour les partitions SlapOS ;
\item Faire une politique en mode \og targeted\fg pour un service déployé ;
\item Automatiser le déploiement des politiques lors de
  l'instanciation du service dans une partition SlapOS ;
\item Créer une politique pour le système SlapOS lui même ;
\item Automatiser l'installation de cet politique au déploiement de SlapOS ;
\item Passer SELinux en mode strict.
\end{enumerate}

\section{Conclusion}\label{trois}

Dans cet article nous avons montré comment \og cloudifier\fg
l'application BOINC dans le cloud SlapOS. Les concepts clés permettant
l'intégration sont les notions de partitions, de stack, de recette.
Bien que conçus initialement pour un usage pour les applications Web,
ils sont suffisamment généraux pour autoriser l'intégration de BOINC à
un coût de développement relativement bas.  Il s'agit avant tout
d'identifier des tournures d'esprit permettant d'intégrer des
composants logiciels communicants.

Le nombre de lignes à écrire pour l'intégration est de l'ordre de 200
et toutes les démarches préopératoires (déclaration de projets,
définition d'un utilisateur\ldots) et postopératoires (re-compilation,
déploiement) sont inchangées par rapport à une application Web.

Pour le futur, nous prévoyons un test d'interopérabilité, par exemple
en prenant des clients dans Azure. Le modèle de déploiement de SlapOS,
o\`u nous pouvons déployer nos propres OS pour le cloud facilite cette
t\^ache. Enfin, des décisions concernant le couplage des politiques
SELinux doivent être prises à partir des pistes que nous avons isolées
dans cet article.

L'objectif général de ce travail est d'offrir des services Desktop
Grid à la demande, sur tous les appareils, sans discernement, c'est à
dire sur les smartphones, tablettes et ordinateurs de bureau. En
outre, le service devrait pouvoir être déployé par un non
informaticien \og en un seul clic\fg et la gestion du système ne devrait
pas être limitée aux administrateurs système, mais largement
ouverte. Il s'agit, de notre point de vue, de conditions nécessaires
pour que les grilles de PC puissent survivre dans les nouveaux
environnements logiciels du Web et soient réellement ouverts,
accessibles au plus grand nombre de chercheurs et d'ingénieurs et non
plus réservés à une communauté d'une dizaine de projets les utilisant.

\nocite{*}
\bibliographystyle{plain}
\bibliography{bib}

\end{document}